\documentclass[prl, twocolumn, showpacs, noshowkeys]{revtex4}
\usepackage{graphics}

\begin{document}
\title{Lorentz transformations that entangle spins \\ and entangle momenta}

\author{Thomas F. Jordan}
\email[email: ]{tjordan@d.umn.edu}
\affiliation{Physics Department, University of Minnesota, Duluth, Minnesota 55812}
\author{Anil Shaji}
\email[email: ]{shaji@unm.edu}
\affiliation{The University of New Mexico, Department of Physics and Astronomy, 800 Yale Blvd. NE, Albuquerque, New Mexico 87131}
\author{E. C. G. Sudarshan}
\email[email: ]{sudarshan@physics.utexas.edu}
\affiliation{The University of Texas at Austin, Center for Statistical Mechanics, 1 University Station C1609, Austin Texas 78712}  

\begin{abstract}
Simple examples are presented of Lorentz transformations that entangle the
spins and momenta of two particles with positive mass and spin 1/2. They
apply to indistinguishable particles, produce maximal entanglement from
finite Lorentz transformations of states for finite momenta, and describe
entanglement of spins produced together with entanglement of momenta. From
the entanglements considered, no sum of entanglements is found to be
unchanged.
\end{abstract}

\pacs{03.30.+p, 03.65.Yz}


\maketitle

The role of relativity in framing statements about quantum information is illustrated by the fact that quantum entanglement can depend on the reference frame of the observer. In particular, Lorentz transformations can change the entanglement of the spins of massive particles \cite{gingrich02a,peres04a}. This happens because Lorentz transformations of the spin of a particle depend on its momentum \cite{czachor97a,peres02a,alsing02a,ahn03a,alsing03a,peres03a,pachos03a,gonera04a,lamata05a,bartlett05a,jordan05a}. There is a rich variety of possibilities. Examples selected for calculations \cite{gingrich02a} have left questions to be answered. Is a separable spin state changed into a maximally entangled state only in the limit of an infinite Lorentz transformation and/or infinite momenta? Does an increase of the entanglement for the spins of two particles require a decrease in the entanglement of their momenta? What happens when the particles are indistinguishable?

Here we give simple examples of Lorentz transformations that entangle the
spins and momenta of two particles with positive mass and spin $1/2$. They
apply to indistinguishable particles, produce maximal entanglement from
finite Lorentz transformations of states for finite momenta, and describe
entanglement of spins produced together with entanglement of momenta. The
operations are made transparent by describing the spin states with density
matrices written in terms of Pauli matrices, so you can see the Pauli
matrices being rotated by the Wigner rotations of the Lorentz
transformations.  

From the entanglements we consider, we find no sum of entanglements that is
unchanged. This leads us to question what is meant by the statements that
``Lorentz boosts introduce a transfer of entanglement between different
degrees of freedom, … . While the entanglement between spin or momentum
alone may change due to Lorentz boosts, the entanglement of the entire wave
function (spin and momentum) is invariant" \cite{gingrich02a} and that
``Entanglement was shown to be an invariant quantity for observers in
uniform motion in the sense that, although different inertial observers may
see these correlations distributed among several degrees of freedom in
different ways, the total amount of entanglement is the same in all
inertial frames" \cite{alsing06a}. We are afraid that these statements
might incorporate an incorrect extrapolation from earlier examples where a
change in the entanglement of the spins was accompanied by a change in the
opposite direction of the entanglement of the momenta \cite{gingrich02a}.

We will consider two particles, $A$ and $B$, that have positive mass and spin $1/2$. We use Pauli matrices $\Sigma_1 , \Sigma_2 , \Sigma_3 $ for the spin of particle $A$, and Pauli matrices $\Xi_1 , \Xi_2 , \Xi_3 $ for the spin of particle $B$. We consider two-particle states described by state vectors or density matrices made from state vectors of the form
\begin{equation}
\label{bp}
| {\bf p}_{A},  {\bf p}_{B} \rangle |{\mbox{spins}} \rangle
\end{equation}
where $|  {\bf p}_{A},   {\bf p}_{B} \rangle  = |  {\bf p}_{A} \rangle_A \: |  {\bf p}_{B} \rangle_B $ is a product state vector of length $1$ that represents a state for the momenta of the two particles composed of a state where the momentum of particle $A$ is concentrated around a value $  {\bf p}_{A} $ and a state where the momentum of particle $B$ is concentrated around a value $  {\bf p}_{B} $, and $|{\mbox{spins}}\rangle $ represents a state for the spins of the two particles. A Lorentz transformation $\Lambda $ changes each $|  {\bf p}\rangle $ to a state vector we call $|  {\bf p} \rangle^\Lambda $ and changes $|  {\bf p}_{A},   {\bf p}_{B} \rangle $ to
\begin{equation}
|  {\bf p}_{A},   {\bf p}_{B} \rangle^\Lambda   = |  {\bf p}_{A} \rangle_A^\Lambda \; |  {\bf p}_{B} \rangle_B^\Lambda   
\end{equation}
which describes momenta concentrated around the Lorentz-transformed values $\Lambda {\bf p}_{A}, \, \Lambda   {\bf p}_{B} $. This is the unitary transformation on the space of momentum states that would represent the Lorentz transformation if the particles had no spins. The Lorentz transformation changes the state vector (\ref{bp}) for momenta and spins to
\begin{equation}
\label{tbp}
|  {\bf p}_{A},   {\bf p}_{B} \rangle^\Lambda  D_A(p_A )D_B(p_B) |{\mbox{spins}} \rangle
\end{equation}
where $D_A $ and $D_B $ are operators on the spin states for particles $A$ and $B$ respectively, $D(p)$ means $D(W(\Lambda ,p))$ with $W(\Lambda , p)$ the Wigner rotation \cite{weinberg95} for the Lorentz transformation $\Lambda $ and the four-vector momentum $p$ corresponding to the three-vector momentum $  {\bf p} $ and the given positive mass, $D_A (W)$ for a rotation $W$ is the $2\times 2$ unitary rotation matrix made from the $\Sigma_j $ so that
\begin{equation}
D_A (W)^\dagger \,  {\bf \Sigma} D_A (W) = W( \bf{\Sigma })
\end{equation}
where $W( \bf{\Sigma })$ is simply the vector $ \bf{\Sigma }$ rotated by $W$, and $D_B(W)$ is the same for $ \bf{\Xi }$. We assume that the momenta are concentrated closely enough that we can use orthogonal state vectors for concentrations around different momentum values, use a single Wigner rotation for each concentration, and accept the accuracy of these approximations.

Suppose that for the state of the pair of particles there are just two pairs of momentum values, $(  {\bf p}_{A_1},   {\bf p}_{B_1})$ and $(  {\bf p}_{A_2},   {\bf p}_{B_2})$, and that the state of the two particles is either a pure state represented by a state vector
\begin{equation}
\label{psv}
\frac{1}{\sqrt{2}} |  {\bf p}_{A_1},   {\bf p}_{B_1} \rangle | 0 \rangle  + \frac{1}{\sqrt{2}} |  {\bf p}_{A_2},   {\bf p}_{B_2} \rangle | 0 \rangle 
\end{equation}
or a mixed state represented by a density matrix
\begin{equation}
\label{msdm}
\frac{1}{2}|  {\bf p}_{A_1},   {\bf p}_{B_1}\rangle \langle   {\bf p}_{A_1},   {\bf p}_{B_1}|\rho  + 
\frac{1}{2}|  {\bf p}_{A_2},   {\bf p}_{B_2}\rangle \langle   {\bf p}_{A_2},   {\bf p}_{B_2}|\rho 
\end{equation}
where
\begin{equation}
\label{R0}
\rho  = |0\rangle \langle 0|
\end{equation}
so that the state of the spins, described by the state vector $|0\rangle $ or the density matrix $\rho $, is the same in both cases.

Let $\rho^\Lambda $ be the density matrix that represents the state of the spins after a Lorentz transformation $\Lambda $. It is obtained by taking the trace over the momentum states $|  {\bf p}_{A},    {\bf p}_{B}\rangle^\Lambda  $ of the density matrix for the state of the two spins after it is changed by $\Lambda $. Each constituent state vector (\ref{bp}) is changed to the spin-rotated state vector (\ref{tbp}). This gives 
\begin{eqnarray}
\rho^\Lambda \! &=& \! \frac{1}{2}  D_A(p_{A_1})D_B(p_{B_1}) \rho D_A(p_{A_1})^\dagger D_B(p_{B_1})^\dagger \nonumber \\
&& + \frac{1}{2} D_A(p_{A_2})D_B(p_{B_2}) \rho D_A(p_{A_2})^\dagger D_B(p_{B_2})^\dagger . 
\end{eqnarray}

Let $ {\bf p}_{A_1} = -  {\bf p}_{B_1} = -  {\bf p}_{A_2} =   {\bf p}_{B_2}$ be along the $x$ axis and let the Lorentz transformation be in the $y$ direction with velocity $v$. Then the Wigner rotations are around the $z$ axis. If $W(\Lambda , p_{A_1})$ and $W(\Lambda , p_{B_2})$ are rotations by $\phi $ around the $z$ axis, then $W(\Lambda , p_{B_1})$ and $W(\Lambda , p_{A_2})$ are rotations by $-\phi $ around the $z$ axis. This can be seen, and $\phi $ can be calculated from $  {\bf p}_{A_1}$ and $v$, by using the formulas from Halpern \cite{halpern68} that we have described \cite[Section V]{jordan05a}. 

Suppose $\rho $ is one of the density matrices
\begin{equation}
\rho _\pm   = \frac{1}{4}[1 \pm  \Sigma _1\Xi _1 \pm  \Sigma _2\Xi _2 - \Sigma _3\Xi _3].
\end{equation}
Both $\rho _+ $ and $\rho _- $ represent maximally entangled pure states for the two spins. They are Bell states. The state of zero total spin is represented by $\rho _- $ and the state obtained from that by rotating one of the spins by $\pi $ around the $z$ axis is represented by $\rho_+ $. The Lorentz transformation takes   $\rho _\pm $ to
\begin{eqnarray}
\label{LTr}
\rho_\pm^\Lambda \!\!\! & = \!\!\!& \frac{1}{2} \Bigl\{  \frac{1}{4} [1  \pm   (\Sigma _1 \cos \phi + \Sigma _2 \sin \phi )(\Xi _1 \cos\phi-\Xi _2 \sin\phi ) \nonumber \\
 & & \quad \pm   (-\Sigma_1 \sin \phi + \Sigma _2 \cos \phi)(\Xi _1 \sin \phi + \Xi_2 \cos \phi)  \nonumber \\
&& \hspace{6.1 cm} - \Sigma_3 \Xi_3] \nonumber \Bigr\} \\
 & & \!\!\!+\frac{1}{2} \Bigl\{ \frac{1}{4}[1  \!\pm \! (\Sigma _1 \cos \phi - \Sigma _2  \sin \phi)(\Xi_1 \cos \phi + \Xi_2 \sin \phi) \nonumber \\
  & & \qquad \pm   \: (\Sigma _1 \sin \phi+ \Sigma _2 \cos \phi)(-\Xi _1 \sin \phi + \Xi _2 \cos \phi)  \nonumber \\ 
&& \hspace{6.1 cm} - \Sigma_3 \Xi_3 ] \Bigr\} \nonumber \\
 & = & \frac{1}{4} [1 \pm (\Sigma_1 \Xi_1 + \Sigma_2 \Xi_2 ) \cos 2 \phi - \Sigma_3 \Xi_3 ]   \nonumber \\
 & = & \rho _\pm \cos^2 \phi \; + \; \rho _\mp \sin^2 \phi .
\end{eqnarray}

We focus on the case where $\phi $ is $\pi /4$. Then the Lorentz transformation takes both $\rho _+ $ and $\rho _- $ to 
\begin{eqnarray}
\label{LTrsimp}
\rho^\Lambda & = & \frac{1}{4}[1 - \Sigma _3 \Xi _3] \nonumber \\
& = & \frac{1}{2}\biggl[ \frac{1}{4} (1-\Sigma_3)(1+\Xi_3) \biggr] \nonumber \\ 
&& \hspace{0 cm}+ \frac{1}{2} \biggl[\frac{1}{4}(1+\Sigma _3) (1-\Xi _3) \!\biggr].
\end{eqnarray}
The Lorentz transformation takes the density matrix $\rho $ for a maximally entangled state to the density matrix $\rho ^\Lambda $ for a separable state that is a mixture of just two products of pure states. The inverse Lorentz transformation of the state of the two particles takes $\rho ^\Lambda $ back to $\rho $.

The Lorentz transformation changes the density matrix $\rho $ of Eq.(\ref{R0}) to
\begin{equation}
\label{R12}
\rho ^\Lambda  = \frac{1}{2}|1\rangle \langle 1| + \frac{1}{2}|2\rangle \langle 2|
\end{equation}
where
\begin{eqnarray}
\label{uuud}
|1\rangle  & = & D_A(p_{A_1})D_B(p_{B_1})|0\rangle \nonumber \\
|2\rangle  & = & D_A(p_{A_2})D_B(p_{B_2})|0\rangle .
\end{eqnarray}
The inner product of these vectors is
\begin{eqnarray}
\label{ip12}
\langle 2|1\rangle & = & \langle 0|[D_A(p_{A_2})]^\dag [D_B(p_{B_2})]^\dag D_A(p_{A_1})D_B(p_{B_1})|0\rangle \vphantom{\bigg|}\nonumber \\  
& = & \langle 0|[D_A(p_{A_1})]^2[D_B(p_{B_1})]^2|0\rangle \vphantom{\bigg|} \nonumber \\
& = & \langle 0|(\cos\phi -i\Sigma _3 \sin\phi  )(\cos\phi +i\Xi _3 \sin\phi  )|0\rangle \vphantom{\bigg|} \nonumber \\
& = & {\mbox{Tr}}[(\cos^2 \phi  -i(\Sigma _3 - \Xi _3 )\cos\phi \sin\phi  \vphantom{\big|} \nonumber \\
&& \hspace{4.1 cm} + \Sigma _3 \Xi _3 \sin^2 \phi )\rho ] \nonumber \\  
& = & \cos^2\phi  - \sin^2\phi  = \cos2\phi \vphantom{\bigg|}
\end{eqnarray}
for the $D_A$, $D_B$ and $\rho $, either $\rho _+ $ or $\rho _- $\,, being considered. When $\phi $ is $\pi /4$, the vectors $|1\rangle $ and $|2\rangle $ are orthogonal, 

The same transformation of the spin density matrix is obtained for different kinds of states for the two particles. If the state of the two particles is a mixture represented by the density matrix (\ref{msdm})
with $\rho $ described by Eq.(\ref{R0}), the Lorentz transformation changes the density matrix to
\begin{eqnarray}
\frac{1}{2} |  {\bf p}_{A_1},   {\bf p}_{B_1}\rangle^{\Lambda \: \Lambda } \langle   {\bf p}_{A_1},   {\bf p}_{B_1}| \otimes |1\rangle \langle 1|  \hspace{3 mm} \nonumber \\
+ \frac{1}{2} |  {\bf p}_{A_2},   {\bf p}_{B_2}\rangle^{\Lambda \: \Lambda } \langle   {\bf p}_{A_2},   {\bf p}_{B_2}| \otimes |2\rangle \langle 2| 
\end{eqnarray}
which for the spins gives the density matrix $\rho ^\Lambda $ described by Eq.(\ref{R12}). If the state of the two particles is a pure state represented by a state vector (\ref{psv}), which for the spins gives the density matrix $\rho $ described by Eq.(\ref{R0}), the Lorentz transformation changes the state vector to 
\begin{equation}
\label{LTsv}
\frac{1}{\sqrt{2}}|  {\bf p}_{A_1},   {\bf p}_{B_1}\rangle^\Lambda  |1\rangle  
 + \frac{1}{\sqrt{2}}|  {\bf p}_{A_2},   {\bf p}_{B_2}\rangle^\Lambda  |2\rangle 
\end{equation}
which for the spins gives the density matrix $\rho ^\Lambda $ described by Eq.(\ref{R12}) again.

The transformation of the density matrix for the momenta is different for the different kinds of states. When the two particles are in a mixed state, the density matrix for the momenta is
\begin{equation}
\label{DP}
\frac{1}{2}|  {\bf p}_{A_1},   {\bf p}_{B_1}\rangle \langle   {\bf p}_{A_1},   {\bf p}_{B_1}|  + 
\frac{1}{2}|  {\bf p}_{A_2},   {\bf p}_{B_2}\rangle \langle   {\bf p}_{A_2},   {\bf p}_{B_2}| 
\end{equation}
before the Lorentz transformation and
\begin{eqnarray}
\label{tDP}
\frac{1}{2}|  {\bf p}_{A_1},   {\bf p}_{B_1}\rangle^{\Lambda \: \Lambda } \langle   {\bf p}_{A_1},   {\bf p}_{B_1}|  \hspace{3 mm} \nonumber \\
 + \frac{1}{2}|  {\bf p}_{A_2},   {\bf p}_{B_2}\rangle^{\Lambda \: \Lambda } \langle   {\bf p}_{A_2},   {\bf p}_{B_2}|
\end{eqnarray}
after the Lorentz transformation. The Lorentz transformation does not change the amount of entanglement in the state of the momenta of the two particles. Both before and after the Lorentz transformation, the state of the momenta is a separable state that is a mixture of just two products of pure states.

When the two particles are in a pure state, the density matrix for the momenta after the Lorentz transformation is described by Eq.(\ref{tDP}), the same as for the mixed state, because the vectors $|1\rangle $ and $|2\rangle $ are orthogonal, for the case where $\phi $ is $\pi /4$ that we are considering. Before the Lorentz transformation, the density matrix for the momenta is that of the pure state represented by the vector
\begin{equation}
\frac{1}{\sqrt{2}}|  {\bf p}_{A_1},   {\bf p}_{B_1}\rangle   
 + \frac{1}{\sqrt{2}}|  {\bf p}_{A_2},   {\bf p}_{B_2}\rangle .  
\end{equation}
For the $  {\bf p}_{A_1} = -  {\bf p}_{B_1} = -  {\bf p}_{A_2} =   {\bf p}_{B_2}$ being considered, this can be chosen to be a maximally entangled state, a Bell state, represented by
\begin{equation}
\label{ud}
\frac{1}{\sqrt{2}}|\alpha  \rangle_A \: |\beta  \rangle_B   
 \pm  \frac{1}{\sqrt{2}}|\beta  \rangle_A \:|\alpha  \rangle_B  
\end{equation}
where $|\alpha \rangle $ represents a state for the momentum of a particle with values concentrated around $  {\bf p}_{A_1} $ and $|\beta  \rangle $ represents a state with momentum values concentrated around $-  {\bf p}_{A_1} $. This choice makes the entanglement of the momentum states the same as that of the spin states. Before the Lorentz transformation, the state described by the vector (\ref{ud}) for the two momenta is maximally entangled. After the Lorentz transformation, the state of the two momenta is represented by the density matrix (\ref{tDP}) with
\begin{eqnarray}
\label{uud}
|  {\bf p}_{A_1},   {\bf p}_{B_1}\rangle^\Lambda  & = & |\alpha  \rangle_A^\Lambda  \; |\beta  \rangle_B^\Lambda  \nonumber \\ 
|  {\bf p}_{A_2},   {\bf p}_{B_2}\rangle^\Lambda  & = & |\beta  \rangle_A^\Lambda  \; |\alpha  \rangle_B^\Lambda  .
\end{eqnarray} 
The state of the momenta is a mixture of two products of pure states, just like the state of the spins described by the density matrix (\ref{LTrsimp}). The Lorentz transformation removes both spin entanglement and momentum entanglement, and the inverse Lorentz transformation restores both. Previous examples \cite{gingrich02a} suggested that a change in the entanglement of the spins might always be accompanied by a change in the opposite direction of the entanglement of the momenta.

The options we have described are flexible enough to apply to indistinguishable particles. The particles are assumed to have spin $1/2$, so if the two particles are indistinguishable, for example if both are electrons, the state of the particles can combine the Bell state for the momenta that is symmetric under interchange of the two particles only with the spin state that is antisymmetric, and the antisymmetric momentum state only with the symmetric spin state.

\begin{figure}
\includegraphics{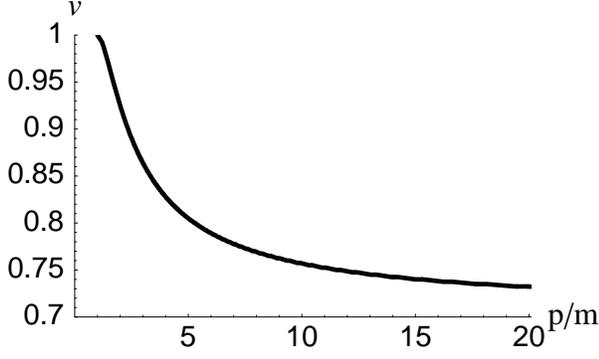}
\caption{Values of the velocity $v$ of the Lorentz transformation and the momentum/mass ratio $|{\bf p}|/m$ that give $\pi/4$ for the angle $\varphi$ of the Wigner rotation.}
\label{fig1}
\end{figure}

Changes between separable states and maximally entangled states are not made here by going to limits. They are made with finite Lorentz transformations of states for finite momenta. Fig. \ref{fig1} shows values of the velocity of the Lorentz transformation and the momentum/mass ratio for the particles that give the $\pi /4$ Wigner rotations that we use. The calculations were made with the formulas from Halpern \cite{halpern68} that we have described \cite[Section V]{jordan05a}. In previous examples \cite{gingrich02a}, maximally entangled states are obtained only in the limit where the velocity of the Lorentz transformation is $c$ and the particle momenta are infinite.

\begin{figure}
\includegraphics{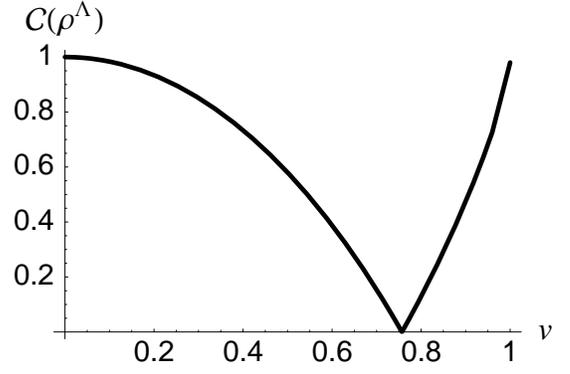}
\caption{The concurrence $C( \rho^{\Lambda})$ as a function of $v$ for $|{\bf p}|/m=10$.}
\label{fig2}
\end{figure}

The change of entanglement is smaller if $\phi $ is not $\pi /4$, but it is still the same for the momenta as for the spins. The state of the spins after the Lorentz transformation is described by one of the density matrices $\rho_\pm^\Lambda $ in Eq.(\ref{LTr}), which we can write as
\begin{equation}
\label{LTr12}
\rho _\pm^\Lambda  = \frac{1}{4} [1 \pm (\Sigma_1 \Xi_1 + \Sigma_2 \Xi_2 ) \cos 2\phi  + (\Sigma_1 \Xi_1 )(\Sigma_2 \Xi_2 ) ]. 
\end{equation}
This shows that for both $\rho_+^\Lambda $ and  $\rho_-^\Lambda $ the eigenvalues are
\begin{equation}
\label{eigenval}
\frac{1}{2} (1+\cos2\phi ), \: \frac{1}{2} (1-\cos2\phi ), \: 0, \: 0 
\end{equation}
because $\Sigma_1 \Xi_1 $ and $\Sigma_2 \Xi_2 $ each have eigenvalues $1$ and $-1$ and together they make a complete set of commuting operators: their four different pairs of eigenvalues label a basis of eigenvectors for the space of states for the two spins. The Wooters concurrence \cite{Wootters98a} is a measure of the entanglement in a state of two qubits. It is defined by
\begin{equation}
\label{eq:note10}
C(\rho ) \equiv {\mbox{max}} \left[ 0, \: \sqrt{\lambda_1} - \sqrt{\lambda_2}-\sqrt{\lambda_3} - \sqrt{\lambda_4} \right]
\end{equation}
where $\rho $ is the density matrix that represents the state and $\lambda_1 , \lambda_2 , \lambda_3 , \lambda_4 $ are the eigenvalues, in decreasing order, of $\rho  \: \Sigma_2 \Xi_2 \: \rho^\star  \:  \Sigma_2 \Xi_2 $, with $\rho^\star $ the complex conjugate that is obtained by changing $\Sigma_2 $ and $\Xi_2 $ to $-\Sigma_2 $ and $-\Xi_2 $. From Eq.(\ref{LTr12}) we have
\begin{equation}
\rho_\pm^\Lambda  \; \Sigma_2 \Xi_2 \; (\rho_\pm^\Lambda )^\star  \; \Sigma_2 \Xi_2  =  \rho_\pm^\Lambda \; (\rho_\pm^\Lambda )^\star \; (\Sigma_2 \Xi_2)^2  = (\rho_\pm^\Lambda )^2 
\end{equation}
so for $\rho_\pm^\Lambda $ the $\sqrt{\lambda_i} $ are the eigenvalues of $\rho_\pm^\Lambda $ and the concurrence is
\begin{equation}
\label{concur}
C(\rho_\pm^\Lambda ) = |\cos2\phi |.
\end{equation}
The concurrence as a function of the velocity of the Lorentz transformation is shown in Fig. \ref{fig2} for the case where $|{\bf p}|/m=10$ for the two particles.

A zero value of the concurrence (\ref{concur}), the mark of a separable state, occurs when $\phi $ is $\pi /4$. We know that a state of two qubits is separable if and only if a positive matrix is the result of taking the partial transpose of the density matrix, the transpose for the states of one of the qubits \cite{peres96a,Horodecki96a}. For $\rho_\pm^\Lambda $, this means changing $\Sigma_2 $ to $-\Sigma_2 $ in the last line of Eq.(\ref{LTr}). The result, with $-\Sigma_3 \Xi_3 $ written as $(\Sigma_1 \Xi_1 )(\Sigma_2 \Xi_2 )$, is
\begin{equation}
\label{partran}
\frac{1}{4} [1 \pm (\Sigma_1 \Xi_1 - \Sigma_2 \Xi_2 ) \cos 2\phi  + (\Sigma_1 \Xi_1 )(\Sigma_2 \Xi_2 ) ]. 
\end{equation}
For either $\rho_+^\Lambda $ or $\rho_-^\Lambda $, this matrix has eigenvalues
\begin{equation}
\label{eigenval2}
\frac{1}{2} \cos 2\phi , \: -\frac{1}{2} \cos 2\phi , \: \frac{1}{2}, \: \frac{1}{2} 
\end{equation}
so the state of the spins after the Lorentz transformation is separable only if $\cos2\phi $ is zero.

We can think of a qubit formed by the two momentum states for each particle and use the concurrence of these qubits for the two particles to measure the entanglement of their momenta. We use Pauli matrices $ \widetilde{\Sigma }_1 $, $ \widetilde{\Sigma }_2$, $ \widetilde{\Sigma }_3$  to describe the momentum qubit for particle $A$ and Pauli matrices $ \widetilde{\Xi  }_1$, $ \widetilde{\Xi  }_2$, $ \widetilde{\Xi  }_3$ for the momentum qubit for particle $B$, taking $|  {\bf p}_{A_1} \rangle $, $|  {\bf p}_{A_2} \rangle $ to be the eigenvectors of $ \widetilde{\Sigma }_3$ for the eigenvalues $1$ and $-1$ and taking $|  {\bf p}_{B_1} \rangle $, $|  {\bf p}_{B_2} \rangle $ to be the eigenvectors of $ \widetilde{\Xi  }_3$. From Eqs.(\ref{LTsv}) and (\ref{ip12}), we find that when the state of the two particles is a pure state, the density matrix for the state of the momenta after the Lorentz transformation is
\begin{eqnarray}
\label{LTtil}
 & & \frac{1}{2}|  {\bf p}_{A_1},   {\bf p}_{B_1}\rangle^{\Lambda \: \Lambda } \langle   {\bf p}_{A_1},   {\bf p}_{B_1}|   \nonumber \\
 && \hspace{5 mm} + \frac{1}{2}|  {\bf p}_{A_2},   {\bf p}_{B_2}\rangle^{\Lambda \: \Lambda } \langle   {\bf p}_{A_2},   {\bf p}_{B_2}| \nonumber \\ 
  & & \hspace{1 cm} + \:  \frac{1}{2}|  {\bf p}_{A_1},   {\bf p}_{B_1}\rangle^{\Lambda \: \Lambda } \langle   {\bf p}_{A_2},   {\bf p}_{B_2}| \cos 2 \phi   \nonumber \\
&&  \hspace{1.5 cm} + \frac{1}{2}|  {\bf p}_{A_2},   {\bf p}_{B_2}\rangle^{\Lambda \: \Lambda } \langle   {\bf p}_{A_1},   {\bf p}_{B_1}| \cos 2 \phi \vphantom{\Biggr\}}\nonumber \\ 
   & = &   \frac{1}{2} \bigg[ \frac{1}{2} (1+ \widetilde{\Sigma }_3 )\frac{1}{2}(1+ \widetilde{\Xi  }_3 )  \nonumber \\
&& \hspace{5 mm} + \frac{1}{2} (1- \widetilde{\Sigma }_3 )\frac{1}{2}(1- \widetilde{\Xi  }_3 ) \nonumber \\
 &  & \hspace{1 cm} + \: \frac{1}{2} ( \widetilde{\Sigma }_1+i \widetilde{\Sigma }_2 )\frac{1}{2} ( \widetilde{\Xi  }_1+i \widetilde{\Xi  }_2 ) \cos 2\phi  \nonumber \\
&& \hspace{1.5 cm} +  \frac{1}{2} ( \widetilde{\Sigma }_1-i \widetilde{\Sigma }_2 )\frac{1}{2} ( \widetilde{\Xi  }_1-i \widetilde{\Xi  }_2 ) \cos2\phi \bigg] \vphantom{\Biggr\}}\nonumber \\
  & = & \frac{1}{4} [1 + ( \widetilde{\Sigma}_1  \widetilde{\Xi}_1 -  \widetilde{\Sigma}_2  \widetilde{\Xi}_2 ) \cos 2\phi  +  \widetilde{\Sigma}_3  \widetilde{\Xi}_3 ] \vphantom{\biggr\}}
\end{eqnarray}
which we can write as
\begin{equation}
\frac{1}{4} [1 + ( \widetilde{\Sigma}_1  \widetilde{\Xi}_1 -  \widetilde{\Sigma}_2  \widetilde{\Xi}_2 ) \cos 2\phi  - ( \widetilde{\Sigma}_1  \widetilde{\Xi}_1 )( \widetilde{\Sigma }_2  \widetilde{\Xi}_2 ) ]. \vphantom{\biggr\}} 
\end{equation}
We can see that this density matrix has the same eigenvalues (\ref{eigenval}) as the density matrices $\rho_\pm^\Lambda $ for the spins, so it has the same Wooters concurrence (\ref{concur}), which is shown in Fig. 2 as a function of the velocity of the Lorentz transformation for the case where. By this measure, the change in entanglement for the momenta is the same as the change in entanglement for the spins.

We can also see that the partial transpose of this density matrix for the momenta, changing $ \widetilde{\Sigma }_2 $ to $- \widetilde{\Sigma }_2 $ in the last line of Eq.(\ref{LTtil}), gives a matrix that has the same eigenvalues (\ref{eigenval2}) as the partial transpose (\ref{partran}) of the density matrix for the spins. For the momenta, as well as for the spins, the state is separable only if $\cos 2 \phi $ is zero.

Both before and after the Lorentz transformation, there is no entanglement or correlation between any momentum and any spin. The density matrix for one momentum and one spin is always the product of a density matrix for the momentum and a density matrix for the spin. There are no two-qubit entanglements or correlations that increase as the entanglement of the spins and the entanglement of the momenta decrease. To see this, we write out the state vector (\ref{LTsv}) as
\begin{eqnarray*}
\frac{1}{\sqrt{2}} | \alpha \rangle_A^{\Lambda} \; | \beta \rangle_B^{\Lambda}  D_A \!\left( p_{A_1} \right) D_B \!\left( p_{B_1} \right) \hspace{1.2 cm} \\
\times \frac{1}{\sqrt{2}}(|+ \rangle_A |- \rangle_B \pm | - \rangle_A |+ \rangle_B)  \\
 \pm \!\frac{1}{\sqrt{2}} \, | \beta \rangle_A^{\Lambda} \; | \alpha \rangle_B^{\Lambda} \, D_A \!\left( p_{A_2} \right) D_B \! \left( p_{B_2} \right) \hspace{1 cm}\\
\times \frac{1}{\sqrt{2}}(|+ \rangle_A |- \rangle_B \pm | - \rangle_A |+ \rangle_B)\\
\end{eqnarray*} 
\vspace{-11 mm}
\begin{eqnarray}
\label{corr1}
=\frac{1}{2} | \alpha \rangle_A^{\Lambda} \; | \beta \rangle_B^{\Lambda} \left( e^{-i \phi} |+ \rangle_A |-\rangle_B \pm e^{i \phi} | - \rangle_A | + \rangle_B \right) \hspace{2 mm}\nonumber \\
\pm  \frac{1}{2} | \beta \rangle_A^{\Lambda} \; | \alpha \rangle_B^{\Lambda} \left( e^{i \phi} |+ \rangle_A |-\rangle_B \pm e^{-i \phi} | - \rangle_A | + \rangle_B \right)
\end{eqnarray}
using Eqs. (\ref{ud}), (\ref{uud}) and (\ref{uuud}) and writing out the Bell-state vector $|0\rangle$ in terms of the eigenvectors $|\pm \rangle_A$ of $\Sigma_3$ and $| \pm \rangle_B$ of $\Xi_3$ for the eigenvalues $\pm 1$. For example, the density matrix for particle $A$, obtained by taking the trace over the states of particle $B$, is
\begin{eqnarray}
\label{corr2}
\rho_A^{\Lambda}& = & \frac{1}{4} \left( |\alpha \rangle_A^{\Lambda} \; \vphantom{\rangle}_A^{\Lambda} \langle \alpha | + |\beta \rangle_A^{\Lambda} \; \vphantom{\rangle}_A^{\Lambda} \langle \beta |  \right) \nonumber \\
&& \hspace{1 cm}\otimes \left( |+ \rangle_A \, \vphantom{\rangle}_A \langle +| + |- \rangle_A \, \vphantom{\rangle}_A \langle - |    \right).
\end{eqnarray}
The density matrix for the momentum of particle $A$ and the spin of particle $B$, obtained by taking the trace over the spin states of $A$ and the momentum states of $B$, is
\begin{eqnarray}
\label{corr3}
\rho_{A_{\mbox{\small momentum}} ,\, B_{\mbox{\small spin}}}^{\Lambda} \!\!\!&=\!\!\!& \frac{1}{4} \left( |\alpha \rangle_A^{\Lambda} \; \vphantom{\rangle}_A^{\Lambda} \langle \alpha | + |\beta \rangle_A^{\Lambda} \; \vphantom{\rangle}_A^{\Lambda} \langle \beta |  \right) \nonumber \\
&& \hspace{0 mm} \otimes \left( |+ \rangle_B \, \vphantom{\rangle}_B \langle +| + |- \rangle_B \, \vphantom{\rangle}_B \langle - |    \right). \quad
\end{eqnarray}
\newpage

\noindent {\bf Acknowledgments: }Anil Shaji acknowledges the support of the US Office of Naval Research Contract No. N000014-03-1-0426.

\vspace{-3 mm}
\bibliography{ncp}

\end{document}